\documentclass[a4paper,twoside,10pt]{article}
\usepackage{amsmath}
\usepackage{graphicx,saj,multicol,subeqnarray,multirow}
\usepackage{longtable}
\usepackage{array}
\usepackage[labelsep=period,labelfont=bf]{caption}

\newcolumntype{C}[1]{>{\centering\let\newline\\\arraybackslash\hspace{0pt}}m{#1}}

\def\papertype{Original Scientific Paper}

\def\udc{...}
\begin{document}
\baselineskip=3.1truemm
\columnsep=.5truecm
\newenvironment{lefteqnarray}{\arraycolsep=0pt\begin{eqnarray}}
{\end{eqnarray}\protect\aftergroup\ignorespaces}
\newenvironment{lefteqnarray*}{\arraycolsep=0pt\begin{eqnarray*}}
{\end{eqnarray*}\protect\aftergroup\ignorespaces}
\newenvironment{leftsubeqnarray}{\arraycolsep=0pt\begin{subeqnarray}}
{\end{subeqnarray}\protect\aftergroup\ignorespaces}
%


\markboth{\eightrm UPDATED RADIO SURFACE-BRIGHTNESS-TO-DIAMETER RELATION
FOR GALACTIC SUPERNOVA REMNANTS} {\eightrm M. Z. PAVLOVI{\'C}, A. DOBARD{\v Z}I{\' C}, B. VUKOTI{\' C} and D. URO{\v S}EVI{\' C}}

{\ }

\publ

\type

{\ }


\title{UPDATED RADIO $\Sigma-D$ RELATION
FOR GALACTIC SUPERNOVA REMNANTS}


\authors{M. Z. Pavlovi{\'c}$^{1}$, A. Dobard{\v z}i{\' c}$^{1}$, B. Vukoti{\' c}$^{2}$ and D. Uro{\v s}evi{\' c}$^{1}$}

\vskip3mm


\address{$^1$Department of Astronomy, Faculty of Mathematics,
University of Belgrade\break Studentski trg 16, 11000 Belgrade,
Serbia}

\Email{marko@math.rs, aleksandra@math.rs, dejanu}{math.rs}

\address{$^2$Astronomical Observatory, Volgina 7, 11060 Belgrade 38, Serbia}

\Email{bvukotic}{aob.rs}


\dates{September 15, 2014}{}


\summary{We present updated empirical radio surface-brightness-to-diameter ($\Sigma-D$) relation for supernova remnants
(SNRs) in our Galaxy. Our original calibration sample of Galactic SNRs with independently determined distances
(Pavlovi{\' c} et al. 2013, hereafter Paper I)
is reconsidered and updated with data which became available in the past two years.
The orthogonal fitting procedure and probability-density-function-based (PDF) method are applied to the calibration sample in the $\log \Sigma - \log D$ plane.
Non-standard orthogonal regression keeps
$\Sigma-D$ and $D-\Sigma$ relations invariant within estimated uncertainties.
Our previous Monte Carlo simulations verified that the slopes of the empirical $\Sigma-D$ relation should
be determined by using orthogonal regression, because of its good performances for data sets with severe scatter.
Updated calibration sample contains 65 shell SNRs. 6 new Galactic SNRs are added to the sample from Paper I,
one is omitted and distances are changed for 10 SNRs. The slope derived here is slightly steeper
($\beta \approx 5.2$) than $\Sigma-D$ slope in Paper I ($\beta \approx 4.8$).
The PDF method relies on data points density maps which can provide more reliable
calibrations that preserve more information contained
in the calibration sample. We estimate distances to five new faint Galactic SNRs discovered for the
first time by Canadian
Galactic Plane Survey and obtained distances of 2.3, 4.0, 1.3, 2.9 and 4.7 kiloparsecs for G108.5+11.0, G128.5+2.6,
G149.5+3.2, G150.8+3.8 and G160.1$-$1.1, respectively. The updated empirical relation is used to estimate
distances of 160 shell Galactic SNRs and new results change their distance scales up to 15 per cent,
compared to results from Paper I. The PDF calculation can provide even few times higher or lower values in
comparison with orthogonal fit, as it uses totally different approach. However, in average, this
difference is 32, 24 and 18 per cent for mode, median and mean distances.
}


\clearpage
\begin{multicols}{2}


\section{1. INTRODUCTION}

The reliable distance determination to Galactic supernova remnants (SNRs) is necessary
for obtaining their basic parameters, such as size, age and explosion energy. It also helps
us to study their evolution and to describe the production of cosmic rays (CRs).
There are several methods for determination of distances to Galactic SNRs such as: from historical
records of supernovae (SNe), from proper motions and radial velocities, kinematic
observations, coincidences with HI, HII and molecular clouds,
OB associations and pulsars, HI absorption and polarization,
optical extinction, low frequency radio absorption, CO emission and from X-ray
observations (Green 1984, Zhu \& Tian 2014).

However, when distance determination with above mentioned methods is not possible,
the distance of a Galactic SNR is commonly estimated
by using the radio surface brightness - diameter relation ($\Sigma-D$).
The relation connecting the radio surface brightness at frequency $\nu$ and diameter of SNR is
given as:
\begin{equation}
\Sigma_{ \nu }(D) = AD^{-\beta},
\end{equation}
\noindent where $A$ is thought to depend on properties of the SN explosion such as the SN explosion energy,
the mass of the ejected matter and also on the properties of ISM such as density of the ISM,
the magnetic field strength, etc, while parameter $\beta$ is thought to be independent of these
properties (Arbutina \& Uro{\v s}evi{\' c} 2005).
Slope $\beta$ explicitly depends on spectral index  $\alpha$ of the integrated radio emission from the remnant
(defined in the sense, $S_{ \nu } \propto \nu^{-\alpha}$, where $S_{ \nu }$ is the flux density
at a frequency $\nu$), as follows from theoretical work (first derived for SNRs; Shklovskii 1960).
This relation is applicable to shell-type SNRs but it can also be
used for composite remnants if we separate surrounding shell flux and flux originating
from the central regions (Case and Bhattacharya 1998).

Despite all the criticism of the $\Sigma-D$ relation (see for example Green 1984, 2004), it
remains an important tool in estimating distances to SNRs in cases where other methods
are not applicable at all.

Calibration is done by linearizing the above Eq. (1) and applying some of the standard
fitting techniques. Parameters $A$ and $\beta$ are then obtained by fitting the data for a sample of SNRs of
known distances (usually called calibrators).
After calibration, the relation can be used to determine distance to a
particular SNR by measuring its flux density and angular diameter. In our previous paper (Pavlovi{\' c} et al. 2013,
hereafter Paper I), we showed that applying of some non-standard fitting procedures,
instead of (standard) vertical regression, can result in different parameters of the $\Sigma-D$ relation
for shell SNRs. We also emphasized very important consequence of our analysis: orthogonal offsets
are more reliable and stable over other types of offsets.

Following our approach from Paper I, we present here updated empirical radio
$\Sigma-D$ relation for Galactic SNRs. New relation is now based on the extended and updated calibration sample,
containing 65 Galactic SNRs: 6 new SNRs were added, one is omitted from previous calibration sample
and distance was changed for 10 remnants in accordance with the new observations.

\section{2. ANALYSIS AND RESULTS}

\subsection{2.1. The updated calibration sample and corresponding $\Sigma-D$ relation}

Updated Galactic sample of 65 shell Galactic SNRs with direct
distance estimates, which is used to derive new $\Sigma-D$ relation,
is in Table 1. The frequency of 1 GHz for surface brightness $\Sigma$
is chosen because flux density measurements at frequencies, presented, both above and
below this value are usually available. Usually, flux density at 1 GHz is not a measured value,
but is derived from the observed radio spectrum of the SNR, following the dependence
$S_{\nu} \propto \nu^{-\alpha}$.

Similar to Paper I, our Galactic sample also includes five composite SNRs for which
it was possible to separate the shell flux density from the pulsar wind nebula
(PWN) flux density or they simply have a pure shell structure in radio domain.
These SNRs are:
G11.2$-$0.3, G93.3+6.9 (DA 530), G189.1+3.0 (IC 443), G338.3$-$0.0 and G344.7$-$0.1.

\end{multicols}

\begin{longtable}{p{0.7cm}llcccp{1cm}}
\caption{Shell SNRs with known distances$^{a}$ - the calibration sample
consists of 65 SNRs. \label{table:Calibrators}}\\

\hline
\hline
 No.  &  Catalog name  & Other name & Surface brightness at 1 GHz &   Distance & Diameter & Ref. \\
        &          &            & ($\times 10^{-21}$ Wm$^{-2}$Hz$^{-1}$sr$^{-1}$)  & (kpc) & (pc) &    \\
\hline
\endfirsthead


\multicolumn{6}{c}{{\tablename} \thetable{} -- Continued} \\[0.5ex]
\hline
 No.  &  Catalog name  & Other name & Surface brightness &   Distance & Diameter & Ref. \\
        &          &            & ($\times 10^{-21}$ Wm$^{-2}$Hz$^{-1}$sr$^{-1}$)  & (kpc) & (pc) &    \\
\hline
\endhead

\hline
\endfoot


\hline
\endlastfoot

1  &  G4.5+6.8$^{b}$ &  Kepler, SN1604, 3C358   &  318	&	6.0 & 5.2  &      1  \\
2  &  G11.2$-$0.3   &       &  193   &   4.4   &   5.1   &                     2  \\
3  &  G18.1$-$0.1    &        & 22.9  &   5.6   &   9.0   &                     3   \\
4  &  G18.8+0.3$^{b}$ & Kes 67 & 26.6   & 14.0   &    55.7         &           4   \\ \smallskip
5  &  G21.8$-$0.6 & Kes 69 & 26    & 5.2  &   30.3     &                      5  \\
6  &  G23.3$-$0.3 & W41 &  14.5  & 4.2         &    33.0      &                6   \\
7  &  G27.4+0.4 & Kes 73, 4C$-$04.71 &  56.4  &   8.65    &   10.1    &         6   \\
8  &  G31.9+0.0$^{b}$ & 3C391  &   103   &   7.2  &    12.4     &             7   \\
9  &  G33.6+0.1$^{b}$  &  Kes 79  & 33.1  & 7.0   & 20.4   &                   8   \\ \smallskip
10 &  G35.6$-$0.4   &               & 8.2 &  3.6  & 13.5  &                    9   \\
11 &  G41.1$-$0.3  & 3C397  &  294  &   10.3  &  10.0  &                       10    \\
12 &  G43.3$-$0.2$^{b}$ & W49B  & 477 & 10.0 &  10.1   &                      11    \\
13 &  G46.8$-$0.3$^{b}$ & HC30  & 9.5  & 7.8   &   33.7  &                    6    \\
14 &  G53.6$-$2.2$^{b}$ & 3C400.2, NRAO 611 & 1.3 &  2.8  &  24.8   &         6       \\ \smallskip
15 &  G54.4$-$0.3$^{b}$  & HC40  &  2.6&   3.3   &   38.4   &               12, 2     \\
16 &  G55.0+0.3 &          &  0.25  &   14.0  &   70.5  &                       6      \\
17 &  G65.1+0.6  &       & 0.18   &   9.0   &  175.6   &                        6      \\
18 &  G74.0$-$8.5$^{b}$  &  Cygnus Loop  &  0.86  &    0.54  & 30.1 &          14      \\
19 &  G78.2+2.1$^{b}$  &   $\gamma$ Cygni, DR4 & 13.4  & 1.20 & 20.9 &         15      \\ \smallskip
20 &  G84.2$-$0.8$^{b}$  &          &   5.2  & 6.0  &  31.2    &              16          \\
21 &  G89.0+4.7$^{b}$  & HB21  &  3.1  & 0.8  &  24.2   &                     6       \\
22 &  G93.3+6.9  &  DA 530, 4C(T)55.38.1  &   2.5    &    2.2  & 14.9  &       37     \\
23 &  G93.7$-$0.2  &  CTB 104A, DA 551  & 1.5  & 1.5  &  34.9  &               6       \\
24 &  G94.0+1.0   &  3C434.1    &  2.6  &   3.0  &  23.9   &                   17        \\ \smallskip
25 &  G96.0+2.0  &        &   0.067  &  4.0   &   30.3   &                       6      \\
26 &  G108.2$-$0.6  &      &  0.32  &  3.2  & 57.2  &                           6      \\
27 &  G109.1$-$1.0  & CTB 109 &  4.2  &  3.2  &  26.1  &                       18         \\
28 &  G111.7$-$2.1$^{b}$  & Cassiopeia A, 3C461 &  16400 & 3.33 & 4.84  &       19           \\
29 &  G114.3+0.3  &       &  0.17 &  0.7  &   14.3  &                           6        \\ \smallskip
30 &  G116.5+1.1$^{b}$  &       &   0.31  &  1.6  &  32.2   &                  6        \\
31 &  G116.9+0.2$^{b}$  &  CTB 1  &   1.04  &  1.6  &  15.8  &                 6        \\
32 &  G119.5+10.2$^{b}$  & CTA 1 &  0.67  &  1.4  &  36.7  &                   6         \\
33 &  G120.1+1.4$^{b}$  & Tycho, 3C10, SN1572  & 132  &  2.5   &  5.8  &      20            \\
34 &  G127.1+0.5  &  R5  &  0.89 &  1.25  & 16.4  &                             6          \\ \smallskip
35 &  G132.7+1.3$^{b}$  & HB3  & 1.06  &  2.2  &  51.2  &                      6         \\
36 &  G152.4$-$2.1  &      & 0.056  &   1.10   &  31.1  &                        21        \\
37 &  G156.2+5.7$^{b}$  &      &  0.062  & 1.0  &      32.0  &                  22         \\
38 &  G160.9+2.6$^{b}$  &  HB9 &  0.98  &   0.8   &     30.2       &           23      \\
39 &  G166.0+4.3$^{b}$  & VRO 42.05.01 &  0.55   &  4.5   &  57.4   &          6     \\ \smallskip
40 &  G180.0$-$1.7 & S147  &   0.3 &   0.62  &  32.5   &                       6      \\
41 &  G189.1+3.0$^{b}$  &  IC443, 3C157 &  12.2  &  1.5  &  19.6  &                   2        \\
42 &  G190.9$-$2.2  &           & 0.047  &   1.0  &  18.8  &                     21       \\
43 &  G205.5+0.5$^{b}$  & Monoceros Nebula &  0.5 &   1.2   & 76.8 &          6       \\
44 &  G260.4$-$3.4$^{b}$ & Puppis A, MSH 08-44  & 6.5 &  2.2 & 35.1  &        6        \\ \smallskip
45 &  G290.1$-$0.8 & MSH 11-61A & 23.8 &  7.0  &   33.2   &                     13        \\
46 &  G292.2$-$0.5 &        &  3.5 &  8.4   &   42.3 &                         6        \\
47 &  G296.5+10.0$^{b}$  &   PKS 1209-51/52 &  1.2  &  2.1  & 46.7 &          24          \\
48 &  G296.7$-$0.9     &             &  3.8 &  9.8   &  31.2  &                25         \\
49 &  G296.8$-$0.3  & 1156-62  &  4.8 &  9.6   &   46.7    &                   6        \\ \smallskip
50 &  G308.4$-$1.4    &          &  2.1   &   9.8   &   12.0   &               26, 27         \\
51 &  G315.4$-$2.3$^{b}$ &  RCW 86, MSH 14-63  &  4.2  & 2.3 & 28.1 &         6         \\
52 &  G327.4+0.4  &  Kes 27  &    10.2      &    4.85  &    29.6   &            6          \\
53 &  G327.6+14.6$^{b}$  &  SN1006, PKS 1459-41  & 3.2 & 1.7 & 14.8  &        28         \\
54 &  G332.4$-$0.4$^{b}$  & RCW 103  &  42.1  &   3.1    & 9.0   &             6          \\ \smallskip
55 &  G337.0$-$0.1  & CTB 33 &  134   &  11.0  &  4.2   &                      6          \\
56 &  G337.8$-$0.1  &  Kes 41  &   50.2  &   11.0  &     23.5  &                6           \\
57 &  G338.3$-$0.0  &          &   15.7  &   11.0  &     25.6  &                29             \\
58 &  G340.6+0.3  &            & 20.9  &    15.0  &   26.2   &                  6          \\
59 &  G344.7$-$0.1  &          &  3.8 &    14.0   &  40.7  &                   30            \\ \smallskip
60 &  G346.6$-$0.2   &          &  18.8   &   7.5  & 17.4 &                     31             \\
61 &  G348.5+0.1$^{b}$  &  CTB 37A  &   48.2  &  9.9   &    43.2    &          32             \\
62 &  G348.7+0.3$^{b}$  &  CTB 37B  &   13.5  &  13.2   &    65.3   &          33              \\
63  & G349.7+0.2$^{b}$  &           &   602  &  11.5   &  7.5  &              34             \\
64  & G352.7$-$0.1  &        &    12.5  &  7.5  &  15.1  &                      8          \\ \smallskip
65 &  G359.1$-$0.5$^{b}$  &        &   3.7   &  7.6  &  53.1  &               35, 36        \\

\hline
\end{longtable}

\noindent
\textbf{Notes.}

\noindent
{$^{a}$ Direct distance estimates, inferred from proper motions, shock and radial velocities,
HI absorption and polarization, association or interaction with HI, HII and CO molecular
clouds, OB associations, pulsars,  X-ray observations, optical extinction and low frequency radio absorption.}

\noindent
{$^{b}$ SNRs from Case \& Bhattacharya's (1998) calibration sample}

\noindent
{\textbf{References}. (1) Chiotellis et al. 2012; (2) Case and Bhattacharya 1998; (3) Leahy et al. 2014; (4) Paron et al. 2012;
 (5) Zhou et al. 2009; (6) Green 2014; Ferrand and Safi-Harb 2012; (7) Su et al. 2014; (8) Giacani et al. 2009;
(9) Zhu et al. 2013; (10) Jiang et al. 2010; (11) Zhu et al. 2014; (12) Junkes et al. 1992; (13) Filipovi{\' c} et al. 2005;
(14) Blair and Sankrit 2005; (15) Uchiyama et al 2002; (16) Leahy and Green 2012; (17) Jeong et al. 2013;
(18) Kothes and Foster 2012; (19) Alarie et al. 2014; (20) Zhang et al. 2013; (21) Foster et al. 2013;
(22) Xu et al. 2007; (23) Leahy and Tian 2007; (24) Giacani et al. 2000; (25) Prinz and Becker 2013;
(26) Prinz and Becker 2012; (27) De Horta et al. 2013; (28) Nikoli{\' c} et al. 2013; (29) Castelletti et al. 2011;
(30) Giacani et al. 2011; (31) Yamauchi et al. 2013; (32) Yamauchi et al. 2014; (33) Tian and Leahy 2012;
(34) Tian and Leahy 2014; (35) Uchida et al. 1992a; (36) Uchida et al. 1992b; (37) Jiang et al. 2007;}

\begin{multicols}{2}

Our main source of information is Green's updated catalog of Galactic SNRs
(2014 May version; Green 2014) and Gilles Ferrand's database of Galactic
SNRs\footnote{A census of high-energy observations of Galactic supernova remnants,
Department of Physics and Astronomy at the University of Manitoba,
www.physics.umanitoba.ca/snr/SNRcat } (Ferrand and Safi-Harb 2012).
We have additionally searched the literature
which provide accurate distances to Galactic shell SNRs
that are not included in above catalogues.

The Galactic sample of calibrators in Paper I contained 60 shell SNRs
with direct distance estimates. We have searched the literature to find recent
and accurate distances to as many SNRs as available. Here presented sample
contains SNRs with revised distances to 10 objects from Paper I sample. SNRs with
new distance estimates are listed in Table 2.

G309.8+0.0 has been removed from the list of calibrators
as we find that its distance questionable. Up to now, the main
reference was Case and Bhattacharya (1998) who proposed the distance
of 3.6 kpc. Actually, these authors cited Huang and Thaddeus (1985)
who further mentioned Caswell et al. (1980) as the main reference for
this distance. Caswell et al. (1980) only
concluded that HI interferometry should permit a distance determination
and they proposed to make this measurement when the improvements to the
Parkes interferometer are completed.

We have also added the following 6 new SNRs
to the original Paper I sample: G18.1$-$0.1, G35.6$-$0.4, G152.4$-$2.1, G190.9$-$2.2,
G296.7$-$0.9 and G308.4$-$1.4.

Foster et al. (2013) reported on the discovery of two Galactic SNRs designated
G152.4$-$2.1 and G190.9$-$2.2, using Canadian Galactic Plane Survey (CGPS) data.
They introduce these two extended faint discrete objects discovered in the CGPS
and show evidence (mainly through their radio spectral and
polarization properties) that classifies them as SNRs.  Foster et al. (2013)
determined systemic local standard of rest (LSR) velocities for both SNRs along their
lines-of-sight using HI and $^{12}$CO$(J=1 \rightarrow 0)$ line data.
They obtained distances of 1.1 $\pm$ 0.1 kpc and 1.0 $\pm$ 0.3 kpc for G152.4−2.1
and G190.9−2.2 respectively.

Two recent studies provided distance estimate to Galactic SNR G18.1$-$0.1 and thus
we include this remnant among calibrators. Paron et al. (2013) suggested that SNR G18.1$-$0.1
is located, along the plane of the sky, close to several HII regions (infrared dust
bubbles N21 and N22, and the HII regions G018.149$-$00.283 and G18.197$-$00.181).
They suggest that all of these objects belong to the same complex at a
distance of about 4 kpc. However, we adopt 5.6 kpc as distance to this SNR
following the conclusions from more recent paper by Leahy et al. (2014). Later, authors analysed radio
and X-ray observations of G18.1$-$0.1 and the overlapping and surrounding HII
regions. The HI spectrum of SNR G18.1$-$0.1 shows absorption up to 100 km s$^{-1}$ but
not beyond, yielding a distance of 5.6 kpc.

\end{multicols}

\begin{longtable}{|l|c|c|c|p{2.8cm}|c|}

\caption{Galactic SNRs with revised distances \label{table:Revised}}\\

\hline
    Catalog name  & Other name & Paper I &   Revised distance & Method & Reference \\
                  &            &        (kpc)        &        (kpc)       &        &           \\
\hline
\endfirsthead


\multicolumn{6}{c}{{\tablename} \thetable{} -- Continued} \\[0.5ex]
\hline
    Catalog name  & Other name & Distance in Paper I &   Revised distance  & Method & Reference \\
                  &            & (kpc)  & (kpc) &    &    \\
\hline
\endhead

\hline
\endfoot


\hline
\endlastfoot

G18.8+0.3  &  Kes 67     & 12.0	&	14.0   &   HI absorption, molecular observations  &     1    \\
G31.9+0.0  &     3C391   &  8.5  &  7.2     &    interaction with molecular clouds  &      2   \\
G84.2-0.8  &         &         4.5        &   6.0      &      HI absorption        &      3        \\
G94.0+1.0  &    3C434.1     &   5.2      &   3.0      &      CO cloud interaction   &        4      \\
G111.7-2.1 &    Cassiopeia A & 3.4   &  3.33 & data from SpIOMM and Hubble\footnote{Authors
used the imaging Fourier transform spectrometer Spectrom{\`e}tre Imageur de l’Observatoire
du Mont-M{\'e}gantic (SpIOMM) to obtain hyperspectral cubes of the Cas A and multi-epoch observations
from the Hubble Space Telescope to create a proper motion map, showing the displacement of several filaments
over the most part of Cas A.} &   5 \\
G120.1+1.4 &	Tycho & 4.0  & 2.5 &     cloud association, HI absorption  &       6     \\
G327.6+14.6	&  SN1006 & 2.2  &  1.7  & shock velocity and proper motion   &      7   \\
G346.6-0.2  &         &  11.0   &  7.5    &   X-ray observations   &       8       \\
G348.5+0.1	&   CTB 37A  &  7.9 &	9.9  &    X-ray observations   &    9   \\
G349.7+0.2  &            &  18.4  &   11.5   &    HI absorption      &   10     \\

\hline
\end{longtable}

\noindent
{\textbf{References}. (1) Paron et al. 2012; Su et al. 2014; (3) Leahy  and Green 2012; (4) Jeong et al. 2013;
(5) Alarie et al. 2014; (6) Zhang et al. 2013; (7) Nikoli{\' c} et al. 2013; (8) Yamauchi et al. 2013;
(9) Yamauchi et al. 2014; (10) Tian and Leahy 2014; }

\begin{multicols}{2}

Prinz and Becker (2013) presented the detailed study of the SNR G296.7$-$0.9 in the 0.2$-$12
keV X-ray band, using data from XMM-Newton. Using the deduced spectral parameters from the
non-equilibrium ionization (NEI) fit, they
derived basic properties of the remnant, such as distance $d$,
post-shock hydrogen density $n_{ \rm{H} }$, swept-up mass $M$, the age of
the remnant $t$, the radius in pc $R_{\rm{s}}$, and the shock velocity $\upsilon_{\rm{s}}$.
Their analysis indicates the SNR with age between 5800 to 7600 years
and a distance of 9.8$^{+1.1}_{-0.7}$ kpc.

Extended radio source in the Galactic plane, G35.6$-$0.4, was reidentified as a SNR by Green (2009) from
radio and infrared survey observations. Zhu et al. (2013) found a plausible distance of 3.6 $\pm$ 0.4 kpc
using HI, $^{13}$CO emission, and HI absorption spectra. With this distance, the average
age of SNR G35.6$-$0.4 would be about 2300 yr and it implies that SNR G35.6$-$0.4 is in an early
evolutionary stage.

Prinz and Becker (2012) presented a detailed X-ray and radio wavelength study of G308.4$-$1.4, a
candidate SNR in the ROSAT All Sky Survey and the MOST SNR
catalog, to identify it as a SNR. The SNR candidate and its central sources were studied
using observations from the Chandra X-ray Observatory, Swift, the Australian Telescope
Compact Array (ATCA) at 1.4 and 2.5 GHz and WISE infrared observation at 24 $\mu$m.
Their analysis revealed that the object is at a distance of 9.8$^{+0.9}_{-0.7}$ kpc and that the
progenitor star exploded 5000 to 7500 years ago. Also, De Horta et al. (2013)
presented radio-continuum observations of this SNR, made with the ATCA,
Molonglo Observatory Synthesis Telescope and the Parkes radio telescope and
confirmed that G308.3$-$1.4 is a SNR with a shell morphology. De Horta et al. (2013)
estimate the flux density at 1 GHz to be $S_{\mathrm{1GHz}} \approx 242$ mJy and
we adopt their value.

\centerline{\includegraphics[
width=1.0\columnwidth, keepaspectratio]{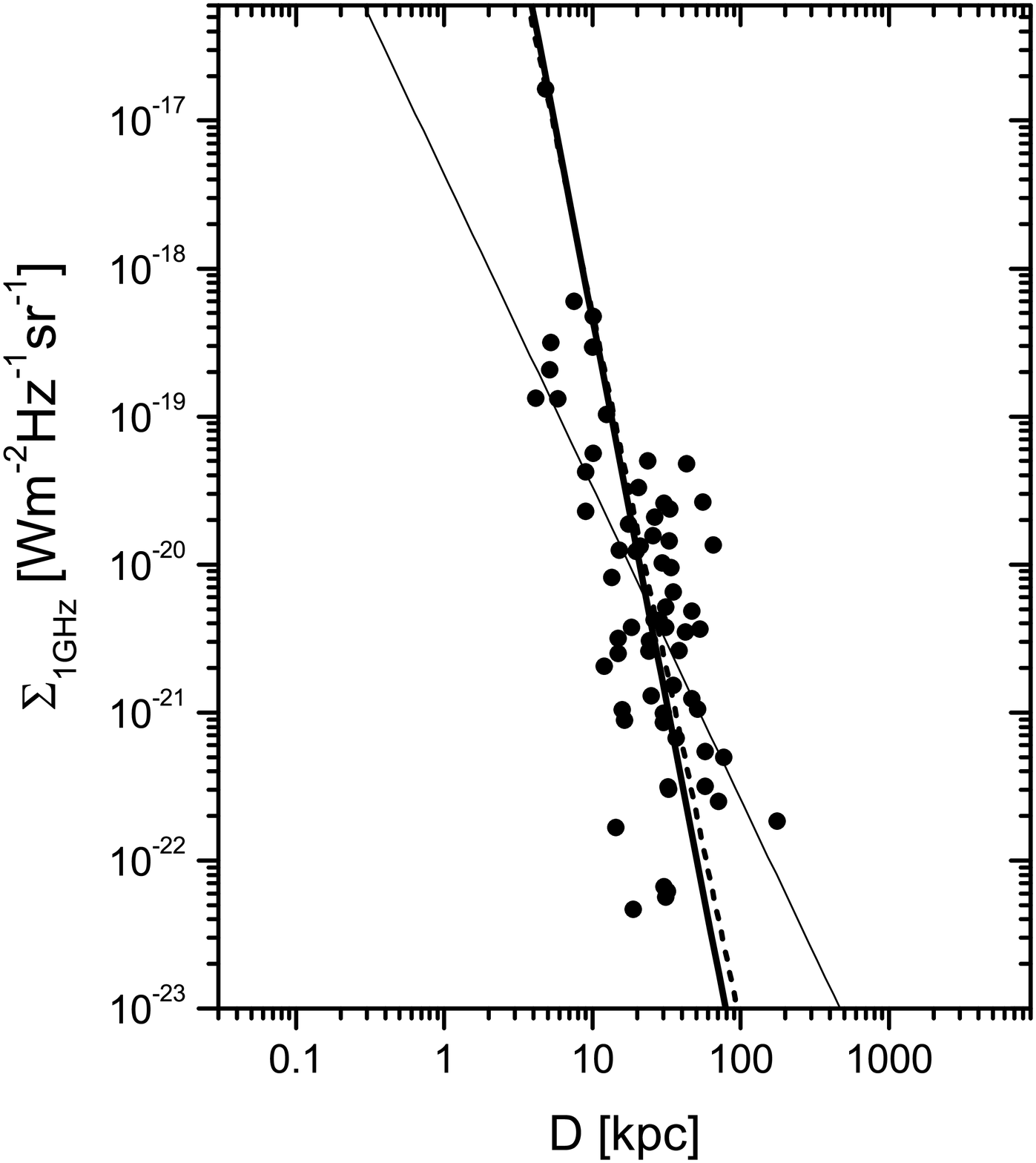}}


\figurecaption{1.}{Surface brightness vs. diameter $\Sigma-D$ relation at 1 GHz for shell
SNRs obtained by using the distance calibrators in Table 1.
The different methods for minimizing the distance of the data from a fitted line are presented.
The thin solid line represent the ordinary (vertical) least squares regression (slope $\beta = 2.1$) while the orthogonal regression
is presented by thick line ($\beta = 5.2$). Dashed line represents orthogonal regression obtained in Paper I ($\beta = 4.8$).
}

SNR G43.3$-$0.2 (W49B) has already been in our previous calibration sample with distance
of 10 kpc. Zhu et al. (2014) confirmed this distance using recent radio and infrared data.
These authors obtained a kinematic distance of $\sim$10 kpc for W49B and suggest that the
SNR is likely associated with the CO cloud.

After applying non-weighted orthogonal regression on the sample containing 65 calibrators
from Table 1, we obtained the relation:
\begin{equation}
\resizebox{0.89\hsize}{!}{$%
\Sigma_{1\mathrm{GHz}} = 6.9^{+460}_{-6.8} \cdot 10^{-14}  D^{-5.2 \pm 1.3} \hspace{0.2cm} \mathrm{W m^{-2} Hz^{-1} sr^{-1}},
$%
}%
\end{equation}

\noindent where asymmetric error interval for parameter $A$ corresponds to $\Delta \log A = 1.83$
($\log A = -13.16$). The new relation is slightly steeper than the one obtained in Paper I ($\beta = 4.8$).
For shell SNRs calculated distances derived from updated $\Sigma-D$ relation are shown in Table 3.
In total, 225 Galactic SNRs are shown: 65 SNRs from our calibration sample (just for comparison) and
160 SNRs without available distance estimates.

In order to obtain an estimate of the accuracy of the obtained $\Sigma-D$ relation
for individual SNR distances, we define fractional errors as:

\begin{equation}
f = \left|  \frac{d_{\mathrm{I}} - d_{\Sigma}}{d_{\mathrm{I}}}  \right|,
\end{equation}

\noindent where $d_{\rm{I}}$ is the independently determined distance to
 an SNR and $d_{\Sigma}$ is the distance derived from our relation. Also,
this is an indicator of the applicability of our relation for distance
determination. The average fractional error for updated sample is $\bar{f}=0.52$ (comparable to
$\bar{f}=0.47$ in Paper I and $\bar{f}=0.41$ obtained by Case and Bhattacharya (1998) for the
significantly smaller calibration sample).

\subsection{2.2. $\Sigma-D$ calibration using PDF-based method}

\centerline{\includegraphics[
width=1.1\columnwidth, keepaspectratio]{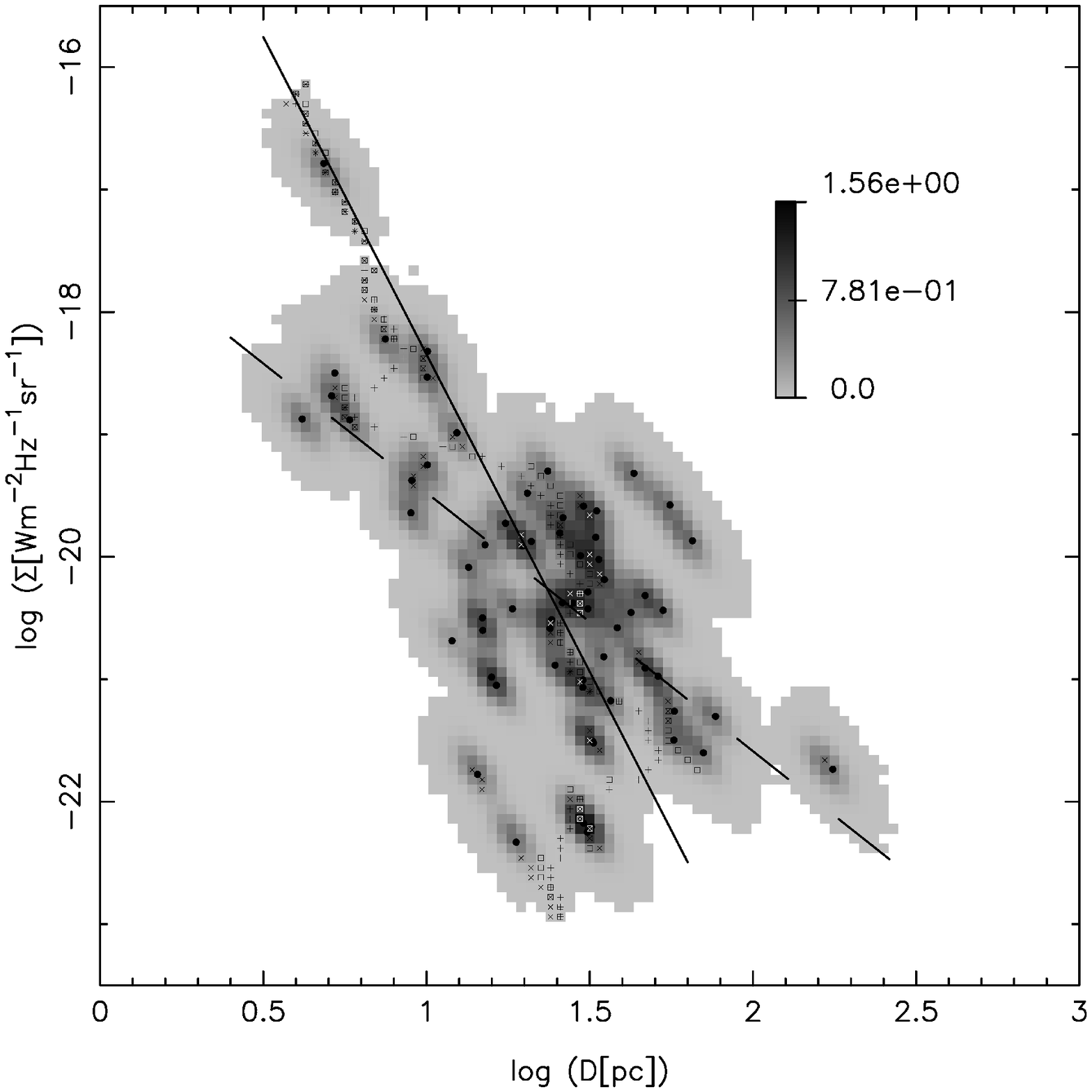}}


\figurecaption{2.}{Greyscale reconstructed data PDF. The lattice of 100 $\times$ 100 cells is mapped on the
variables range shown in the plot.
The markers represent parameters of the distributions at fixed $\Sigma$ values, along the $D$ axis
(rows of the plotted PDF matrix): mode – diagonal cross, median – open square and mean – cross.
The dashed line represents the orthogonal offsets best-fitting line and solid line represents
ordinary least-squares regression.
}

Calibration of the $\Sigma-D$ relation can also be done without using orthogonal fitting nor any
other standard fitting procedure.
Instead, as pointed out by Vukoti{\' c} et al. (2014), random resampling can be used to obtain the
probability density function (PDF) of calibration data points in the fitting plane. Therefore,
the resulting PDF is used to estimate distance-related properties. Vukoti{\' c} et al. (2014) showed that
PDF-based method for calibration can provide more accurate and more reliable calculations
than those obtained by standard linear fitting procedures. Detailed description of algorithm for the
calculation of data points density distribution is given in Vukoti{\' c} et al. (2014).

In our analysis, $10^{6}$ random resamplings have been done and we mapped resulting samples on the
$10^2 \times 10^2$ lattice spanning the coordinates range shown in Fig. 2.
After applying the algorithm from Vukoti{\' c} et al. (2014), the resulting data sample PDF is obtained in the
form of a 2D matrix that can be used as the pattern for distance determination (Fig. 2). Thus, this PDF matrix
contains more information about the calibration sample than just the line of the best fit.
In order to obtain a single value for the distance to a particular SNR, we get a value of the diameter $D$
from the corresponding PDF distribution of $D$ at the particular fixed
value of $\log\Sigma$. This PDF distribution of $D$ for fixed value of $\log\Sigma$ actually represents
1-dimensional "slice" from 2-dimensional PDF matrix obtained for the entire data sample.

In Table 3 we present distances to 225 Galactic SNRs inferred from PDF calibration, in form of three
basic statistical properties of these distributions: median, mode and mean. The median represents the distance
corresponding to diameter $D$ with equal probability that the value of $D$ is situated in higher or
lower values than median and it changes very slowly with data fluctuations. The mode represents the
value which corresponds to diameter $D$ having the highest probability and it is representative
in cases where this mode peak dominates the entire distribution. The mean distance can be useful in
estimating error, although it is more sensitive to fluctuations in data than median value
(Vukoti{\' c} et al. 2014).

\subsection{2.3. Distances to five newly detected Galactic SNRs}

Gerbrandt et al. (2014) presented the results of a systematic search of the Canadian Galactic Plane Survey
(CGPS) for faint, extended non-thermal structures that are likely shells of uncatalogued Galactic SNRs.
They discovered five new objects which are strong candidates for new SNRs.
These five objects are designated by their Galactic coordinate names G108.5+11.0, G128.5+2.6, G149.5+3.2, G150.8+3.8 and
G160.1$-$1.1. CGPS 1420 MHz polarization data and 4.8 GHz polarization data also provide evidence
that these objects are newly discovered SNRs.

Gerbrandt et al. (2014) estimated flux densities at 4.8 GHz, 2.7 GHz, 1420 MHz, 408 MHz, and 327 MHz
for each object except G108.5+11.0, for which only three frequency measurements were available: 4.85 GHz,
1420 MHz, and 408 MHz. They also provide flux densities at 1 GHz, deduced by fitting
a power-law $S_{\nu} \propto \nu^{-\alpha}$ to the frequency spectrum. We then applied our
updated empirical Galactic $\Sigma-D$ relation as well as PDF-based calibration
to obtain the diameters and hance distance estimates for these five objects. Flux densities
at 1 GHz, angular dimensions and obtained distances are given in Table 3.

\end{multicols}

\begin{longtable}{|l|c|c|C{2.4cm}|c|c|c|C{2cm}|}

\caption{Distances to five newly discovered candidates for SNRs, along with their integrated flux and
spectral properties. \label{table:NewSNRs}}\\

\hline
      \multirow{2}{*}{Catalog name}   &   \multirow{2}{*}{Flux density}  &   \multirow{2}{*}{Angular size}  &  \multirow{2}{*}{Orthogonal fit}   &  \multicolumn{3}{c|}{PDF distance} &
 \multirow{2}{*}{Adopted}  \\ [+1.0ex]
                  & $S_{1\mathrm{GHz}}$ (mJy)  &   (arcmin)    &  distance  &  \multicolumn{3}{c|}{(kpc)} & distance\\ \cline{5-7}
                  &        &       &   (kpc)    & Mode   &  Median   &    Mean &   (kpc)  \\

\hline
\endfirsthead


\multicolumn{6}{c}{{\tablename} \thetable{} -- Continued} \\[0.5ex]
\hline
    SNR name  &  Flux density  & Angular size &  Orthogonal fit distance  &  \multicolumn{3}{c|}{PDF distance} \\
\hline
\endhead

\hline
\endfoot


\hline
\endlastfoot

G108.5+11.0  &    734     &   64.9 $\times$ 39.0  &	   4.1   &   2.3  &   2.2  &   1.8 & 2.3 (1.3)  \\
G128.5+2.6   &    255     &   39.6 $\times$ 21.5  &	   7.0   &   4.0  &   3.7  &   3.0 & 4.0 (2.3)  \\
G149.5+3.2   &    590     &   55.6 $\times$ 49.3  &	   4.1   &   1.3  &   1.5  &   1.7 & 1.3 (2.2)  \\
G150.8+3.8   &    665     &   64.1 $\times$ 18.8  &	   5.2   &   2.9  &   2.9  &   2.7 & 2.9   \\
G160.1$-$1.1  &    265     &  35.9 $\times$ 13.2  &	   8.3   &   4.7  &   4.7  &   4.3 & 4.7   \\

\hline
\end{longtable}

\begin{multicols}{2}

We also give the PDF of the diameter variable at the fixed value of $\log\Sigma$
(respectively -22.3601, -22.3461, -22.4895, -22.0807, -22.0749) for these five
very low surface brightness objects (Fig. 3). As noted by Vukoti{\' c} et al. (2014),
in case when the distance estimates for mode, mean and median are
close together, then mode value should be used as the most probable
one. In other cases, where difference is significant, an
inspection of PDF may be required, either from a data sample
PDF (Fig. 3), or directly from 2D matrix of PDF presented in form of graph
(Fig. 2). We adopted mode distances for these five new SNRs because
all three estimators (mode, median and mean) were within the range of the highest
peak (Fig. 3). As can be seen from Fig. 3, PDF distributions for SNRs G108.5+11.0, G128.5+2.6
and G149.5+3.2 have two dominant peaks of approximately the same probability, so we adopt two
distance estimates for them (last column values in brackets correspond to lower peaks).

The results presented in Table 3 leads to a conclusion that
PDF-based method gives lower diameters than the values estimated from the
best fit line (orthogonal fit) and therefore gives lower distances for
a given angular diameter of the object. The explanation for this can be
a denser populated calibrator data point region at $\log\Sigma \approx -22$ which
is situated to the left of the orthogonally fitted line, towards the lower diameters.

These values are obtained using the calibrators from Table 1,
however give a significantly different results than by using the orthogonal fitting.
The average fractional errors (defined by Eq. (3)) for PDF calibration
are 0.35, 0.43 and 0.39 for mode, median and mean
distances respectively, and they are notably lower than that in
orthogonal fitting ($\bar{f}=0.52$). Thus, the PDF method ensures greater
consistency and more accurate calibrations.

\end{multicols}

\centerline{\includegraphics[
width=0.95\columnwidth, keepaspectratio]{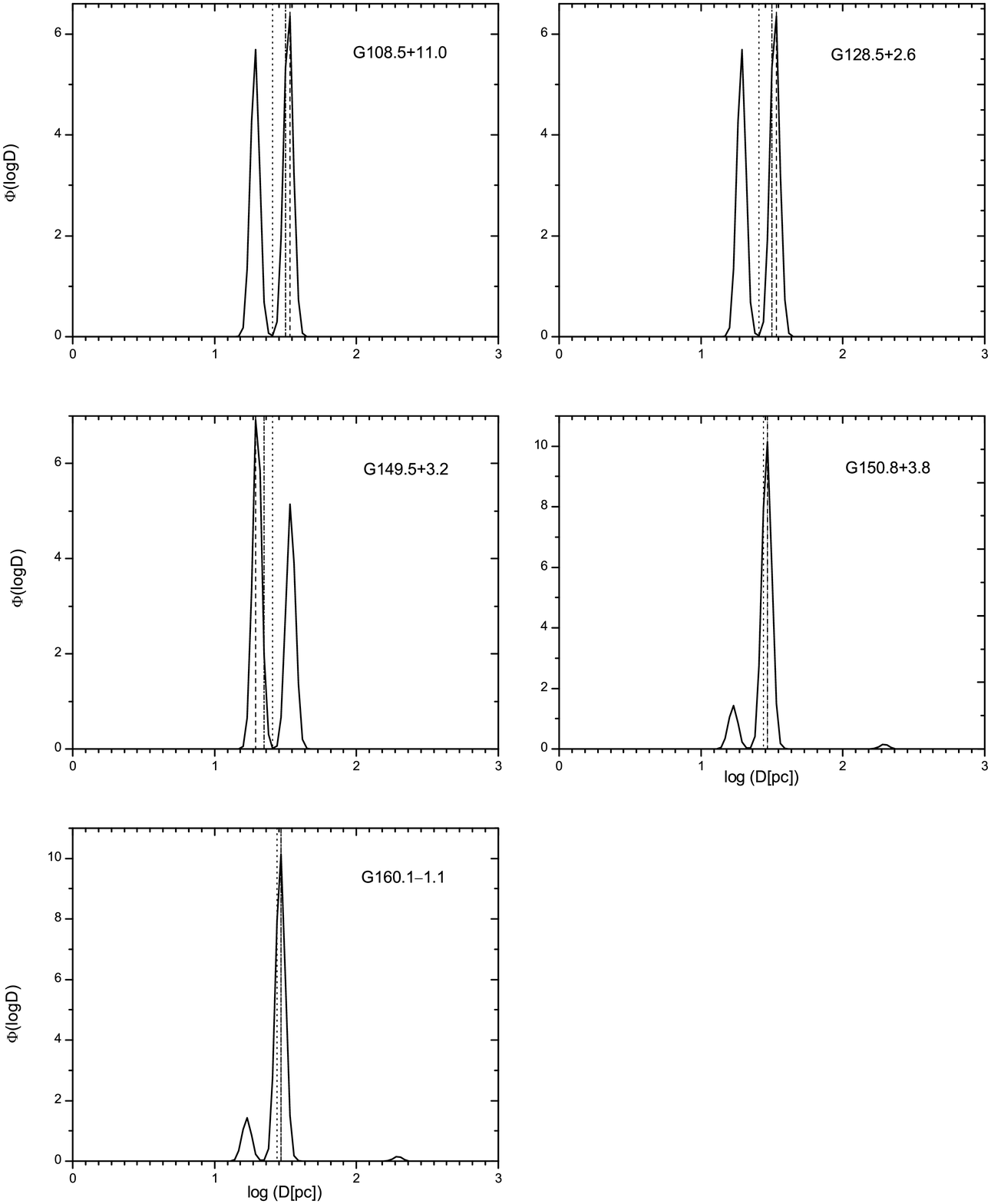}}

\textbf{Fig. 3.} PDF of the diameter variable at the fixed value of $\log\Sigma$
from the data sample. PDFs are given respectively from G108.5+11.0 (top left)
until G160.1$-$1.1 (bottom left). The first three SNRs have two peaks of similar
probability, so we put lower peak values in brackets in the last column of Table 3.
Mode, median and mean are presented with dashed, dash–dotted and dotted lines,
respectively. The PDF function is such that $\int^{+\infty}_{-\infty}\Phi(\log D)\mathrm{d}(\log D)=1$ (Vukotic et al. 2014).

\begin{multicols}{2}

\section{3. DISCUSSION}

The $\Sigma-D$ relation is an important tool in estimating distances to Galactic SNRs in cases where other
distance measurements are not applicable. However, we acknowledge
its theoretical and statistical inconsistencies. Also, caution is necessary
as uncertainties of distance estimates could be higher than 50 \% for values obtained
using orthogonal fitting and at least 35 \% for PDF-based distance calculation.

The catalog of known Galactic SNRs has grown since our Paper I, from 274 to 294 SNRs.
The total number of SNRs with known distances (calibrators) has increased and
our Galactic sample now contains 65 shell remnants, which is used to derive new empirical $\Sigma-D$ relation.
Our improved sample contains revised distances to 10 SNRs from previous sample
and we have also added 6 new SNRs to the previous sample.
One calibrator is omitted from previous calibration sample.
In addition, we also provide a PDF-based calibration for the updated sample.

Our obtained slope
($\beta = 5.2$) is slightly steeper than our slope in Paper I ($\beta = 4.8$),
but it stays within the statistical errors. These two $\Sigma-D$ lines intersect in point close
to $D \approx 7.9$ pc and $\Sigma \approx 1.6 \times 10^{-17}$ $\mathrm{Wm^{-2}Hz^{-1}sr^{-1}}$
and therefore updated relation gives lower distances than those in Paper I for SNRs with surface brightness
$\Sigma<1.6 \times 10^{-17}$ $\mathrm{Wm^{-2}Hz^{-1}sr^{-1}}$ (almost all Galactic SNRs, exception among
calibrators is Cassiopeia A). This difference increases linearly with decreasing radio surface brightness.
Results obtained with updated relation can change the distance scale for Galactic SNRs up to about 15 per cent,
in comparison with Paper I. Obtained slope is still far from the trivial one ($\beta = 2.0$, Uro{\v s}evi{\' c} 2002, 2003), and it
agrees with theoretical predictions for the Sedov phase of SNR evolution (Sedov subphases should have
$\beta$ slopes from 2 to 5.75, Paper I).

Significant number of selection effects impose different limitations on statistical studies of Galactic SNRs
(e.g., Green 1991; Uro{\v s}evi{\'c} et al. 2005, 2010). The identification of Galactic SNRs is always
accompanied by the difficulty in identifying (1) faint SNRs and (2) small angular size SNRs.
The surface-brightness limit affects the completeness of all surveys of Galactic
SNRs and the flux densities of many SNRs are poorly determined.
Furthermore, Uro{\v s}evi{\' c} et al. (2010) concluded that the sensitivity selection effect
does not have a major impact on the $\Sigma-D$ slope for the sample of SNRs from starburst
galaxy M82. Additionally, Malmquist bias is present in the Galactic samples, making them
incomplete. This is a volume selection effect that favors bright objects because they are sampled from
a larger spatial volume in any flux-limited surveys and therefore, it acts against low
surface-brightness remnants. Only an extragalactic set of SNRs does not suffer from Malmquist
bias which is distance dependent selection effect.

It has been generally accepted that no single $\Sigma-D$ relation can be constructed for
entire known sample of Galactic SNRs (Arbutina and Uro{\v s}evi{\' c} 2005).
Our current calibration sample contains 65 SNRs in total, which probably have different
explosion energies, evolve in different ambient density, and may be in different phases
of SNR evolution. Consequently, any single linear $\Sigma-D$ relation represents only an
averaged evolutionary track for Galactic SNRs. Nevertheless, we argue that orthogonal
fitting instead of ordinary least-squares regression (vertical fitting) represents a
significant step forward and improves the statistical analysis of the $\Sigma-D$ relation.

PDF calibration, however, represents another improvement of the statistical
approach to $\Sigma-D$ relation.
It is undoubtedly more reliable method than just a single fitting calibration
(orthogonal fitting in our case), as it gets a firmer hold on very important SNRs complex
evolutionary features. It should be also emphasised the importance of local PDF features in
statistical distance calibrations that are averaged out in the case of orthogonal fitting.
PDF-based statistics is one of the few approaches that can probably
deal with different classes of SNRs which is essential for gaining better understanding
of SNR evolution.

Nevertheless, orthogonal fitting stays an important method. Having best fitting line
parameters, one can simply calculate the distance to an SNRs without
using specialized bootstrap code which constructs the PDF distribution for a given
sample of calibrators. Also, obtained slope of the empirical $\Sigma-D$ relation, may have
important consequences for the theoretical modeling of SNR evolution.

For a proper $\Sigma-D$ analysis, the $L-D$ correlation should
be also tested to allow a possible dependence of luminosity on the
diameter as $L_{\nu} = CD^{\delta}$ , where $C$ is constant, following
from the luminosity-surface-brightness relation $L_{\nu} = \pi^2 D^2 \Sigma_{\nu}$
(Arbutina et al. 2004). However, our Monte Carlo simulations in Paper I have revealed
that the $L-D$ relation is very sensitive to scatter in the data. Therefore,
it is not possible to obtain the $L-D$ relation for our sample as it is
subject to severe scatter and obtained slopes does not have physical meaning.

\section{4. SUMMARY}

We present a re-analysis of the empirical Galactic $\Sigma-D$ relation for Galactic SNRs.
Empirical $\Sigma-D$ relation is strongly dependent not only on regression type due
to severe data scatter, but also on the calibration sample containing SNRs with independently
determined distances. Following the main conclusion from Paper I, that the orthogonal regression
is the most accurate slope predictor in data sets with severe scatter, we derive updated relation
only by using orthogonal regression. We have not analyzed possible
dependence of radio luminosity on the linear diameter ($L-D$ relation) because
this relation is even more sensitive to severe data scatter.

We applied our updated empirical relation to estimate distances to 160 shell-like remnants with
unknown distances (Table 5 contains 225 SNRs in total, calibrators are also included).
We also give distances inferred from PDF calibration, in the form of three basic statistical
properties of these distributions: median, mode and mean.

We have also applied our relation for estimating distances to five new faint SNRs, discovered with CGPS.
PDF-based calculation gives the following results for these SNRs: 2.3 or 1.3 kpc for G108.5+11.0, 4.0 or
2.3 kpc for G128.5+2.6, 1.3 or 2.2 kpc for G149.5+3.2 kpc, 2.9 kpc G150.8+3.8 and 4.7 kpc for G160.1$-$1.1.
Two distance estimates for each of SNRs G108.5+11.0, G128.5+2.6 and G149.5+3.2 have approximately the same
probability but nevertheless, we suggest using mode values 2.3, 4.0 and 1.3 kiloparsecs corresponding to
the most dominant peak in PDF distribution. For SNRs G150.8+3.8
and G160.1$-$1.1 the most probable value (mode) can be used with high confidence.

Although we applied two different methods for distance determination, we expect results obtained
by using PDF-based statistics to be more accurate and more reliable. Nevertheless, PDF-based
estimates should also be used with caution (as well as those obtained with orthogonal fitting)
because uncertainties could be as large as about 35-40 \% (Vukoti{\' c} et al. 2014).

Taking into account a typical evolution timescale ($10^5$ yr) of
SNRs and an event rate of two supernovae per century
in the Galaxy (Dragicevich et al. 1999), 2000 SNRs are expected in our Galaxy.
Thus, there are still many "missing" Galactic SNRs,
due to difficulties in identifying low surface brightness
objects as well as due to non-conspicuous unresolved "point"-like appearance of very distant SNRs.
In the future, high resolution and sensitivity large-scale radio, X$-$ray and $\gamma-$ray surveys of our Galaxy would be crucial in the detection of never-before-seen faint SNRs.


\acknowledgements{During the work on this paper, the authors were financially
supported by the Ministry of Education and Science of the Republic
of Serbia through the Project No. 176005 'Emission nebulae: structure and evolution'.
B.V. also acknowledges financial support through the Project No. 176021 'Visible and invisible
matter in nearby galaxies: theory and observations'.
}

\newpage

\end{multicols}

\begin{longtable}{|l|c|c|C{1.5cm}|C{1.5cm}|c|c|c|}

\caption{Distances to 225 shell SNRs calculated from updated $\Sigma-D$ relation and PDF-based analysis. \label{table:Calculated-update}}\\

\hline

\multirow{2}{*}{Catalog name}  &  \multirow{2}{*}{Other name} &    \multirow{2}{*}{Flux density}  &   \multicolumn{2}{c|}{Orthogonal fit}   & \multicolumn{3}{c|}{PDF distance} \\ [+1.0ex]
         &              &  (Jy)   &   \multicolumn{2}{c|}{  } &  \multicolumn{3}{c|}{(kpc)}\\  \cline{4-8}
       &                &       &   Diameter (pc)  &  Distance (kpc)    & Mode   &  Median   &    Mean  \\

\hline
\endfirsthead


\multicolumn{5}{c}{{\tablename} \thetable{} -- Continued} \\[0.5ex]
\hline

\multirow{2}{*}{Catalog name}  &  \multirow{2}{*}{Other name} &    \multirow{2}{*}{Flux density}  &   \multicolumn{2}{c|}{Orthogonal fit}   & \multicolumn{3}{c|}{PDF distance} \\ [+1.0ex]
         &              &  (Jy)   &   \multicolumn{2}{c|}{  } &  \multicolumn{3}{c|}{(kpc)}\\  \cline{4-8}
       &                &       &   Diameter (pc)  &  Distance (kpc)    & Mode   &  Median   &    Mean  \\

\hline
\endhead

\hline
\endfoot


\hline
\endlastfoot

G0.0+0.0   &     Sgr A East
    &     100.0       &       7.7       &       9.0  &  7.5    &     7.5    &   7.5  \\
G0.3+0.0    &
    &     22.0       &       17.1       &       5.4  &  9.3    &     8.1    &   7.5  \\
G1.0-0.1    &
    &     15.0       &       16.3       &       7.0  &  3.9    &     10.3    &   9.0  \\
G1.4-0.1    &
    &     2.0       &       26.3       &       9.0  &  8.2    &     8.8    &   8.8  \\
G3.7-0.2    &
    &     2.3       &       27.8       &       7.7  &  6.6    &     7.1    &   7.1  \\
G3.8+0.3    &
    &     3.0       &       30.5       &       5.8  &  8.5    &     5.6    &   5.3  \\
G4.2-3.5    &
    &     3.2       &       35.7       &       4.4  &  6.7    &     4.8    &   4.8  \\
G4.5+6.8$^{a}$     &     Kepler, SN1604, 3C358
    &     19.0       &       10.7       &       12.3 &  11.2    &     11.2    &   9.1  \\
G4.8+6.2    &
    &     3.0       &       30.5       &       5.8  &  8.5    &     5.6    &   5.3  \\
G5.2-2.6    &
    &     2.6       &       31.3       &       6.0  &  5.3    &     5.6    &   5.3  \\
G5.5+0.3    &
    &     5.5       &       24.2       &       6.2  &  7.1    &     7.6    &   7.6  \\
G5.9+3.1    &
    &     3.3       &       31.2       &       5.4  &  4.7    &     5.1    &   4.7  \\
G6.1+0.5    &
    &     4.5       &       26.1       &       6.1  &  5.6    &     6.0    &   6.0  \\
G6.4+4.0    &
    &     1.3       &       44.2       &       4.9  &  18.4    &     7.0    &   5.7  \\
G6.5-0.4    &
    &     27.0       &       20.0       &       3.8  &  3.7    &     5.3    &   4.9  \\
G7.0-0.1    &
    &     2.5       &       29.4       &       6.7  &  10.2    &     6.3    &   6.3  \\
G7.2+0.2    &
    &     2.8       &       26.4       &       7.6  &  6.9    &     7.4    &   7.4  \\
G7.7-3.7    &     1814-24
    &     11.0       &       25.6       &       4.0  &  4.6    &     4.6    &   4.3  \\
G8.3-0.0    &
    &     1.2       &       21.3       &       16.3  &  24.3    &     22.7    &   19.8  \\
G8.7-5.0    &
    &     4.4       &       32.6       &       4.3  &  3.9    &     4.2    &   3.9  \\
G8.7-0.1    &     (W30)
    &     80.0       &       23.0       &       1.8  &  2.6    &     2.4    &   2.3  \\
G8.9+0.4    &
    &     9.0       &       27.6       &       3.9  &  3.4    &     3.7    &   3.7  \\
G9.7-0.0    &
    &     3.7       &       25.7       &       6.9  &  7.9    &     7.9    &   7.4  \\
G9.8+0.6    &
    &     3.9       &       24.8       &       7.1  &  8.5    &     8.5    &   7.9  \\
G9.9-0.8    &
    &     6.7       &       22.3       &       6.4  &  9.7    &     9.1    &   7.9  \\
G10.5-0.0    &
    &     0.9       &       25.2       &       14.4  &  16.9    &     16.9    &   15.8  \\
G11.0-0.0    &
    &     1.3       &       28.5       &       9.8  &  8.3    &     8.9    &   8.9  \\
G11.1-1.0    &
    &     5.8       &       24.8       &       5.8  &  6.9    &     6.9    &   6.4  \\
G11.1-0.7    &
    &     1.0       &       28.6       &       11.2  &  9.4    &     10.1    &   10.1  \\
G11.1+0.1    &
    &     2.3       &       26.5       &       8.3  &  7.5    &     8.1    &   8.1  \\
G11.2-0.3$^{a}$    &
    &     22.0       &       11.6       &       10.0  &  4.5    &     4.8    &   5.2  \\
G11.4-0.1    &
    &     6.0       &       19.5       &       8.4  &  8.4    &     11.0    &   11.0  \\
G11.8-0.2    &
    &     0.7       &       22.6       &       19.4  &  29.1    &     27.2    &   25.4  \\
G12.2+0.3    &
    &     0.8       &       24.9       &       15.6  &  18.5    &     18.5    &   17.3  \\
G12.7-0.0    &
    &     0.8       &       25.7       &       14.7  &  16.9    &     16.9    &   15.8  \\
G13.5+0.2    &
    &     3.5       &       17.3       &       13.3  &  22.7    &     19.8    &   18.4  \\
G14.1-0.1    &
    &     0.5       &       27.2       &       17.1  &  15.1    &     16.1    &   16.1  \\
G14.3+0.1    &
    &     0.6       &       24.3       &       18.7  &  22.7    &     22.7    &   21.2  \\
G15.1-1.6    &
    &     5.5       &       31.6       &       4.1  &  3.5    &     3.8    &   3.5  \\
G15.4+0.1    &
    &     5.6       &       24.9       &       5.9  &  7.0    &     7.0    &   6.5  \\
G15.9+0.2    &
    &     5.0       &       18.0       &       10.4  &  18.4    &     14.9    &   13.9  \\
G16.0-0.5    &
    &     2.7       &       26.8       &       7.5  &  6.7    &     7.2    &   7.2  \\
G16.2-2.7    &
    &     2.5       &       30.9       &       6.2  &  9.0    &     6.0    &   5.6  \\
G16.4-0.5    &
    &     4.6       &       24.8       &       6.5  &  7.8    &     7.8    &   7.3  \\
G17.0-0.0    &
    &     0.5       &       26.3       &       18.1  &  16.5    &     17.7    &   17.7  \\
G17.4-2.3    &
    &     5.0       &       30.9       &       4.4  &  6.4    &     4.2    &   3.9  \\
G17.4-0.1    &
    &     0.4       &       29.4       &       16.9  &  25.6    &     15.8    &   15.8  \\
G17.8-2.6    &
    &     5.0       &       30.9       &       4.4  &  6.4    &     4.2    &   3.9  \\
G18.1-0.1$^{a}$    &
    &     4.6       &       20.5       &       8.8  &  13.6    &     11.8    &   11.0  \\
G18.6-0.2    &
    &     1.4       &       23.1       &       13.2  &  19.4    &     18.1    &   16.9  \\
G18.8+0.3$^{a}$     &     Kes 67
    &     33.0       &       17.3       &       4.3  &  7.4    &     6.5    &   6.0  \\
G19.1+0.2    &
    &     10.0       &       28.3       &       3.6  &  3.1    &     3.3    &   3.3  \\
G20.4+0.1    &
    &     9.0       &       18.0       &       7.8  &  13.6    &     11.0    &   10.3  \\
G21.0-0.4    &
    &     1.1       &       27.0       &       11.7  &  10.4    &     11.1    &   11.1  \\
G21.5-0.1    &
    &     0.4       &       27.4       &       18.9  &  16.5    &     17.7    &   17.7  \\
G21.6-0.8    &
    &     1.4       &       31.1       &       8.2  &  7.3    &     7.8    &   7.3  \\
G21.8-0.6$^{a}$     &     Kes 69
    &     65.0       &       17.5       &       3.0  &  5.1    &     4.4    &   4.1  \\
G22.7-0.2    &
    &     33.0       &       22.1       &       2.9  &  4.5    &     4.2    &   3.6  \\
G23.3-0.3$^{a}$     &     W41
    &     70.0       &       19.4       &       2.5  &  2.5    &     3.3    &   3.3  \\
G24.7-0.6    &
    &     8.0       &       23.5       &       5.4  &  6.3    &     6.8    &   6.8  \\
G25.1-2.3    &
    &     8.0       &       37.1       &       2.6  &  3.9    &     3.9    &   3.1  \\
G27.4+0.0$^{a}$     &     4C-04.71
    &     6.0       &       14.9       &       12.8  &  8.4    &     18.0    &   14.6  \\
G28.6-0.1    &
    &     3.0       &       25.0       &       8.0  &  9.4    &     9.4    &   8.8  \\
G29.6+0.1    &
    &     1.5       &       21.3       &       14.6  &  21.7    &     20.3    &   17.7  \\
G30.7+1.0    &
    &     6.0       &       28.2       &       4.7  &  4.0    &     4.3    &   4.3  \\
G31.5-0.6    &
    &     2.0       &       33.0       &       6.3  &  5.6    &     6.0    &   5.6  \\
G31.9+0.0$^{a}$     &     3C391
    &     25.0       &       13.2       &       7.7  &  3.5    &     3.5    &   4.0  \\
G32.0-4.9    &     3C396.1
    &     22.0       &       33.0       &       1.9  &  1.7    &     1.8    &   1.7  \\
G32.4+0.1    &
    &     0.2       &       32.2       &       18.5  &  16.9    &     18.1    &   16.9  \\
G32.8-0.1    &     Kes 78
    &     11.0       &       23.2       &       4.7  &  6.9    &     6.4    &   6.0  \\
G33.2-0.6    &
    &     3.5       &       29.6       &       5.7  &  8.5    &     5.3    &   5.3  \\
G33.6+0.1$^{a}$     &     Kes 79, 4C00.70, HC13
    &     20.0       &       16.9       &       5.8  &  10.1    &     8.8    &   7.7  \\
G35.6-0.4$^{a}$     &
    &     9.0       &       21.6       &       5.8  &  8.5    &     7.9    &   6.9  \\
G36.6-0.7    &
    &     1.0       &       42.8       &       5.9  &  4.7    &     8.1    &   7.1  \\
G36.6+2.6    &
    &     0.7       &       37.5       &       8.7  &  12.7    &     12.7    &   11.1  \\
G40.5-0.5    &
    &     11.0       &       25.6       &       4.0  &  4.6    &     4.6    &   4.3  \\
G41.1-0.3$^{a}$     &     3C397
    &     25.0       &       10.6       &       10.9  &  10.0    &     10.0    &   8.1  \\
G41.5+0.4    &
    &     1.0       &       30.0       &       10.3  &  15.4    &     10.1    &   9.5  \\
G42.0-0.1    &
    &     0.5       &       31.5       &       13.5  &  11.8    &     12.7    &   11.8  \\
G42.8+0.6    &
    &     3.0       &       34.1       &       4.9  &  4.5    &     4.9    &   4.5  \\
G43.3-0.2$^{a}$     &     W49B
    &     38.0       &       9.9       &       9.8  &  9.7    &     9.1    &   8.4  \\
G43.9+1.6    &
    &     9.0       &       39.2       &       2.2  &  1.8    &     3.1    &   2.7  \\
G45.7-0.4    &
    &     4.2       &       30.9       &       4.8  &  7.0    &     4.6    &   4.3  \\
G46.8-0.3$^{a}$     &     (HC30)
    &     17.0       &       20.3       &       4.7  &  4.5    &     6.4    &   5.9  \\
G49.2-0.7    &     (W51)
    &     160.0       &       17.2       &       2.0  &  3.4    &     2.9    &   2.7  \\
G53.6-2.2$^{a}$     &     3C400.2, NRAO 611
    &     8.0       &       30.9       &       3.5  &  5.1    &     3.3    &   3.1  \\
G54.4-0.3$^{a}$     &     (HC40)
    &     28.0       &       27.0       &       2.3  &  2.1    &     2.2    &   2.2  \\
G55.0+0.3$^{a}$     &
    &     0.5       &       42.4       &       8.4  &  6.7    &     11.7    &   10.2  \\
G55.7+3.4    &
    &     1.0       &       41.4       &       6.2  &  5.1    &     8.8    &   7.7  \\
G57.2+0.8    &     (4C21.53)
    &     1.8       &       28.8       &       8.2  &  6.9    &     7.4    &   7.4  \\
G59.5+0.1    &
    &     3.0       &       28.4       &       6.5  &  5.5    &     5.9    &   5.9  \\
G64.5+0.9    &
    &     0.2       &       39.7       &       17.1  &  13.6    &     23.6    &   20.6  \\
G65.1+0.6$^{a}$     &
    &     5.5       &       45.0       &       2.3  &  0.7    &     3.5    &   2.5  \\
G65.3+5.7    &
    &     42.0       &       52.3       &       0.7  &  0.4    &     0.4    &   0.3  \\
G67.7+1.8    &
    &     1.0       &       33.6       &       8.6  &  8.1    &     8.7    &   8.1  \\
G69.7+1.0    &
    &     2.0       &       30.7       &       7.1  &  10.3    &     6.8    &   6.3  \\
G73.9+0.9    &
    &     9.0       &       28.8       &       3.7  &  3.1    &     3.3    &   3.3  \\
G74.0-8.5$^{a}$     &     Cygnus Loop
    &     210.0       &       33.5       &       0.6  &  0.6    &     0.6    &   0.6  \\
G78.2+2.1$^{a}$     &     DR4, gamma Cygni SNR
    &     320.0       &       19.7       &       1.1  &  1.1    &     1.6    &   1.5  \\
G82.2+5.3    &     W63
    &     120.0       &       26.4       &       1.2  &  1.0    &     1.1    &   1.1  \\
G83.0-0.3    &
    &     1.0       &       27.5       &       11.9  &  10.4    &     11.1    &   11.1  \\
G84.2-0.8$^{a}$     &
    &     11.0       &       23.7       &       4.5  &  5.3    &     5.7    &   5.7  \\
G89.0+4.7$^{a}$     &     HB21
    &     220.0       &       26.2       &       0.9  &  0.8    &     0.9    &   0.9  \\
G93.3+6.9$^{a}$     &     DA 530, 4C(T)55.38.1
    &     9.0       &       27.2       &       4.0  &  3.5    &     3.8    &   3.8  \\
G93.7-0.2$^{a}$     &     CTB 104A, DA 551
    &     65.0       &       29.9       &       1.3  &  1.9    &     1.2    &   1.2  \\
G94.0+1.0$^{a}$     &     3C434.1
    &     13.0       &       27.0       &       3.4  &  3.0    &     3.2    &   3.2  \\
G96.0+2.0$^{a}$     &
    &     0.3       &       53.2       &       7.0  &  3.9    &     3.9    &   3.6  \\
G108.2-0.6$^{a}$     &
    &     8.0       &       40.5       &       2.3  &  1.8    &     3.1    &   2.7  \\
G109.1-1.0$^{a}$     &     CTB 109
    &     22.0       &       24.6       &       3.0  &  3.6    &     3.6    &   3.4  \\
G111.7-2.1$^{a}$     &     Cassiopeia A, 3C461
    &     2720.0       &       5.0       &       3.4  &  3.4    &     3.4    &   3.4  \\
G114.3+0.3$^{a}$     &
    &     5.5       &       45.9       &       2.2  &  0.7    &     3.3    &   2.3  \\
G116.5+1.1$^{a}$     &
    &     10.0       &       40.6       &       2.0  &  1.6    &     2.7    &   2.4  \\
G116.9+0.2$^{a}$     &     CTB 1
    &     8.0       &       32.2       &       3.3  &  3.0    &     3.2    &   3.0  \\
G119.5+10.2$^{a}$     &     CTA 1
    &     36.0       &       35.1       &       1.3  &  2.1    &     1.5    &   1.5  \\
G120.1+1.4$^{a}$     &     Tycho, 3C10, SN1572
    &     56.0       &       12.7       &       5.4  &  2.4    &     2.4    &   2.6  \\
G126.2+1.6    &
    &     6.0       &       45.0       &       2.2  &  0.7    &     3.3    &   2.4  \\
G127.1+0.5$^{a}$     &     R5
    &     12.0       &       33.2       &       2.5  &  2.3    &     2.4    &   2.3  \\
G132.7+1.3$^{a}$     &     HB3
    &     45.0       &       32.1       &       1.4  &  1.2    &     1.3    &   1.2  \\
G152.4-2.1$^{a}$     &
    &     3.5       &       56.6       &       2.0  &  1.1    &     1.1    &   1.0  \\
G156.2+5.7$^{a}$     &
    &     5.0       &       55.5       &       1.7  &  1.0    &     1.0    &   0.9  \\
G160.9+2.6$^{a}$     &     HB9
    &     110.0       &       32.6       &       0.9  &  0.8    &     0.8    &   0.8  \\
G166.0+4.3$^{a}$     &     VRO 42.05.01
    &     7.0       &       36.5       &       2.9  &  4.3    &     4.3    &   3.5  \\
G178.2-4.2    &
    &     2.0       &       54.7       &       2.8  &  1.5    &     1.5    &   1.4  \\
G179.0+2.6    &
    &     7.0       &       43.7       &       2.1  &  8.2    &     3.1    &   2.5  \\
G180.0-1.7$^{a}$    &     S147
    &     65.0       &       40.9       &       0.8  &  0.6    &     1.0    &   0.9  \\
G182.4+4.3    &
    &     0.5       &       63.9       &       4.4  &  1.4    &     1.5    &   1.6  \\
G189.1+3.0$^{a}$     &     IC443, 3C157
    &     164.7       &       20.0       &       1.5  &  1.5    &     2.1    &   2.0  \\
G190.9-2.2$^{a}$     &
    &     1.3       &       58.7       &       3.1  &  1.7    &     1.7    &   1.4  \\
G192.8-1.1    &     PKS 0607+17
    &     20.0       &       37.2       &       1.6  &  2.4    &     2.4    &   2.1  \\
G205.5+0.5$^{a}$     &     Monoceros Nebula
    &     140.0       &       38.1       &       0.6  &  0.9    &     0.9    &   0.7  \\
G206.9+2.3    &     PKS 0646+06
    &     6.0       &       39.2       &       2.8  &  2.2    &     3.9    &   3.4  \\
G213.0-0.6    &
    &     21.0       &       47.4       &       1.1  &  0.3    &     0.8    &   1.0  \\
G260.4-3.4$^{a}$     &     Puppis A, MSH 08-44
    &     130.0       &       22.6       &       1.4  &  2.1    &     2.0    &   1.9  \\
G261.9+5.5    &
    &     10.0       &       31.1       &       3.1  &  2.7    &     2.9    &   2.7  \\
G266.2-1.2    &     RX J0852.0-4622
    &     50.0       &       36.8       &       1.1  &  1.6    &     1.6    &   1.3  \\
G272.2-3.2    &
    &     0.4       &       41.9       &       9.6  &  7.8    &     13.5    &   11.8  \\
G279.0+1.1    &
    &     30.0       &       37.1       &       1.3  &  2.0    &     2.0    &   1.7  \\
G284.3-1.8    &     MSH 10-53
    &     11.0       &       26.5       &       3.8  &  3.4    &     3.7    &   3.7  \\
G286.5-1.2    &
    &     1.4       &       30.7       &       8.4  &  12.3    &     8.1    &   7.6  \\
G289.7-0.3    &
    &     6.2       &       25.2       &       5.5  &  6.4    &     6.4    &   6.0  \\
G290.1-0.8$^{a}$     &     MSH 11-61A
    &     42.0       &       17.6       &       3.7  &  6.7    &     5.4    &   5.1  \\
G292.2-0.5$^{a}$     &
    &     7.0       &       25.5       &       5.1  &  5.9    &     5.9    &   5.5  \\
G294.1-0.0    &
    &     2.0       &       44.8       &       3.9  &  1.2    &     5.8    &   4.1  \\
G296.1-0.5    &
    &     8.0       &       30.9       &       3.5  &  5.0    &     3.3    &   3.1  \\
G296.5+10.0$^{a}$     &     PKS 1209-51/52
    &     48.0       &       31.2       &       1.4  &  1.2    &     1.3    &   1.2  \\
G296.7-0.9$^{a}$     &
    &     3.0       &       25.2       &       7.9  &  9.3    &     9.3    &   8.6  \\
G296.8-0.3$^{a}$     &     1156-62
    &     9.0       &       24.0       &       4.9  &  5.7    &     6.1    &   6.1  \\
G298.6-0.0    &
    &     5.0       &       22.3       &       7.4  &  11.2    &     10.5    &   9.1  \\
G299.2-2.9    &
    &     0.5       &       39.2       &       9.6  &  7.7    &     13.4    &   11.7  \\
G299.6-0.5    &
    &     1.0       &       33.2       &       8.8  &  7.8    &     8.4    &   7.8  \\
G301.4-1.0    &
    &     2.1       &       39.3       &       4.6  &  3.7    &     6.5    &   5.6  \\
G302.3+0.7    &
    &     5.0       &       27.0       &       5.5  &  4.9    &     5.2    &   5.2  \\
G304.6+0.1    &     Kes 17
    &     14.0       &       16.6       &       7.1  &  12.7    &     11.0    &   9.6  \\
G306.3-0.9    &
    &     0.2       &       30.0       &       25.8  &  38.4    &     25.4    &   23.7  \\
G308.1-0.7    &
    &     1.2       &       32.1       &       8.5  &  7.3    &     7.8    &   7.3  \\
G308.4-1.4$^{a}$     &
    &     0.2       &       37.1       &       15.0  &  22.3    &     22.3    &   18.1  \\
G309.2-0.6    &
    &     7.0       &       23.1       &       5.9  &  8.7    &     8.1    &   7.6  \\
G309.8+0.0    &
    &     17.0       &       23.5       &       3.7  &  4.3    &     4.7    &   4.7  \\
G310.6-0.3    &     Kes 20B
    &     5.0       &       20.2       &       8.7  &  8.4    &     11.8    &   11.0  \\
G310.8-0.4    &     Kes 20A
    &     6.0       &       22.8       &       6.5  &  9.7    &     9.1    &   8.5  \\
G311.5-0.3    &
    &     3.0       &       18.6       &       12.8  &  17.7    &     17.7    &   16.5  \\
G312.4-0.4    &
    &     45.0       &       24.1       &       2.2  &  2.5    &     2.7    &   2.7  \\
G312.5-3.0    &
    &     3.5       &       30.2       &       5.5  &  8.1    &     5.3    &   5.0  \\
G315.4-2.3$^{a}$     &     RCW 86, MSH 14-63
    &     49.0       &       24.7       &       2.0  &  2.4    &     2.4    &   2.3  \\
G315.9-0.0    &
    &     0.8       &       39.9       &       7.3  &  5.8    &     10.1    &   8.8  \\
G316.3-0.0    &     (MSH 14-57)
    &     20.0       &       22.1       &       3.8  &  5.8    &     5.4    &   4.7  \\
G317.3-0.2    &
    &     4.7       &       23.1       &       7.2  &  10.6    &     9.9    &   9.2  \\
G318.2+0.1    &
    &     3.9       &       38.4       &       3.5  &  5.0    &     5.0    &   4.4  \\
G321.9-1.1    &
    &     3.4       &       35.3       &       4.3  &  6.7    &     4.8    &   4.8  \\
G321.9-0.3    &
    &     13.0       &       26.7       &       3.4  &  3.1    &     3.3    &   3.3  \\
G323.5+0.1    &
    &     3.0       &       26.9       &       7.1  &  6.3    &     6.8    &   6.8  \\
G327.2-0.1    &
    &     0.4       &       27.4       &       18.9  &  16.5    &     17.7    &   17.7  \\
G327.4+0.4$^{a}$     &     Kes 27
    &     30.0       &       20.7       &       3.4  &  5.2    &     4.5    &   4.2  \\
G327.4+1.0    &
    &     1.9       &       30.2       &       7.4  &  11.0    &     7.2    &   6.8  \\
G327.6+14.6$^{a}$     &     SN1006, PKS 1459-41
    &     19.0       &       26.0       &       3.0  &  3.4    &     3.4    &   3.2  \\
G329.7+0.4    &
    &     34.0       &       25.0       &       2.4  &  2.8    &     2.8    &   2.6  \\
G330.0+15.0    &     Lupus Loop
    &     350.0       &       29.6       &       0.6  &  0.9    &     0.5    &   0.5  \\
G330.2+1.0    &
    &     5.0       &       22.8       &       7.1  &  10.6    &     9.9    &   9.2  \\
G332.0+0.2    &
    &     8.0       &       21.6       &       6.2  &  9.1    &     8.5    &   7.4  \\
G332.4-0.4$^{a}$     &     RCW 103
    &     28.0       &       15.8       &       5.4  &  3.1    &     7.7    &   6.7  \\
G332.4+0.1    &     MSH 16-51, Kes 32
    &     26.0       &       18.7       &       4.3  &  5.9    &     5.9    &   5.5  \\
G332.5-5.6    &
    &     2.0       &       42.6       &       4.2  &  3.3    &     5.8    &   5.0  \\
G335.2+0.1    &
    &     16.0       &       23.4       &       3.8  &  4.5    &     4.8    &   4.8  \\
G336.7+0.5    &
    &     6.0       &       22.7       &       6.6  &  9.8    &     9.2    &   8.6  \\
G337.0-0.1$^{a}$     &     (CTB 33)
    &     1.5       &       13.4       &       30.6  &  27.6    &     20.9    &   19.5  \\
G337.2-0.7    &
    &     1.5       &       22.8       &       13.1  &  19.4    &     18.1    &   16.9  \\
G337.3+1.0    &     Kes 40
    &     16.0       &       19.7       &       5.0  &  5.0    &     7.1    &   6.6  \\
G337.8-0.1$^{a}$     &     Kes 41
    &     18.0       &       15.3       &       7.1  &  4.6    &     9.8    &   7.9  \\
G338.1+0.4    &
    &     4.0       &       26.9       &       6.2  &  5.5    &     5.9    &   5.9  \\
G338.3-0.0$^{a}$     &
    &     6.7       &       19.1       &       8.2  &  8.4    &     11.0    &   11.0  \\
G340.4+0.4    &
    &     5.0       &       20.6       &       8.4  &  13.0    &     11.3    &   10.6  \\
G340.6+0.3$^{a}$     &
    &     5.0       &       18.1       &       10.4  &  18.1    &     14.7    &   13.7  \\
G341.9-0.3    &
    &     2.5       &       21.9       &       10.8  &  16.6    &     15.5    &   13.5  \\
G342.0-0.2    &
    &     3.5       &       23.9       &       7.9  &  9.1    &     9.8    &   9.8  \\
G342.1+0.9    &
    &     0.5       &       33.6       &       12.2  &  11.5    &     12.3    &   11.5  \\
G343.1-0.7    &
    &     7.8       &       28.2       &       4.1  &  3.5    &     3.7    &   3.7  \\
G344.7-0.1$^{a}$     &
    &     2.5       &       25.2       &       8.7  &  10.1    &     10.1    &   9.5  \\
G345.7-0.2    &
    &     0.6       &       27.2       &       15.6  &  13.7    &     14.7    &   14.7  \\
G346.6-0.2$^{a}$     &
    &     8.0       &       18.4       &       7.9  &  11.0    &     11.0    &   10.3  \\
G347.3-0.5    &     RX J1713.7-3946
    &     30.0       &       31.1       &       1.8  &  2.6    &     1.7    &   1.6  \\
G348.5-0.0    &
    &     10.0       &       19.3       &       6.6  &  6.7    &     8.8    &   8.8  \\
G348.5+0.1$^{a}$     &     CTB 37A
    &     72.0       &       15.4       &       3.5  &  2.1    &     5.1    &   4.5  \\
G348.7+0.3$^{a}$     &     CTB 37B
    &     26.0       &       19.7       &       4.0  &  3.9    &     5.6    &   5.2  \\
G349.2-0.1    &
    &     1.4       &       25.0       &       11.7  &  13.8    &     13.8    &   12.9  \\
G349.7+0.2$^{a}$     &
    &     20.0       &       9.5       &       14.5  &  11.4    &     12.2    &   12.2  \\
G350.0-2.0    &
    &     26.0       &       28.6       &       2.2  &  1.8    &     2.0    &   2.0  \\
G351.7+0.8    &
    &     10.0       &       23.0       &       5.0  &  7.3    &     6.8    &   6.4  \\
G351.9-0.9    &
    &     1.8       &       27.2       &       9.0  &  7.9    &     8.5    &   8.5  \\
G352.7-0.1$^{a}$     &
    &     4.0       &       20.0       &       9.9  &  9.7    &     13.7    &   12.8  \\
G353.6-0.7    &
    &     2.5       &       38.4       &       4.4  &  6.3    &     6.3    &   5.5  \\
G353.9-2.0    &
    &     1.0       &       33.2       &       8.8  &  7.8    &     8.4    &   7.8  \\
G354.8-0.8    &
    &     2.8       &       31.5       &       5.7  &  5.0    &     5.3    &   5.0  \\
G355.4+0.7    &
    &     5.0       &       31.4       &       4.3  &  3.8    &     4.1    &   3.8  \\
G355.6-0.0    &
    &     3.0       &       21.1       &       10.5  &  15.7    &     14.6    &   12.8  \\
G355.9-2.5    &
    &     8.0       &       22.2       &       5.9  &  9.0    &     8.4    &   7.3  \\
G356.2+4.5    &
    &     4.0       &       32.7       &       4.5  &  4.1    &     4.3    &   4.1  \\
G356.3-1.5    &
    &     3.0       &       30.0       &       6.0  &  8.9    &     5.9    &   5.5  \\
G356.3-0.3    &
    &     3.0       &       23.1       &       9.1  &  13.3    &     12.4    &   11.6  \\
G357.7+0.3    &
    &     10.0       &       27.0       &       3.9  &  3.4    &     3.7    &   3.7  \\
G358.0+3.8    &
    &     1.5       &       46.5       &       4.2  &  1.3    &     3.3    &   4.0  \\
G358.1+0.1    &
    &     2.0       &       34.3       &       5.9  &  5.4    &     5.8    &   5.4  \\
G358.5-0.9    &
    &     4.0       &       28.2       &       5.7  &  4.9    &     5.2    &   5.2  \\
G359.0-0.9    &
    &     23.0       &       22.6       &       3.4  &  5.1    &     4.7    &   4.4  \\
G359.1-0.5$^{a}$     &
    &     14.0       &       25.3       &       3.6  &  4.2    &     4.2    &   3.9  \\
G359.1+0.9    &                                   &     2.0       &       27.7       &
  8.3  &  7.2    &     7.7    &   7.7  \\

\hline

\end{longtable}
\noindent
{\textbf{Notes}.

\noindent $^a$ SNRs belonging to our updated calibration sample from Table 1.
As their distances have been calculated by using orthogonal fit or PDF-based
method, these distances could be significantly different than those
in Table 1 which were obtained by using methods mentioned in Paragraph 1.}

\begin{multicols}{2}


\references

Alarie, A., Bilodeau, A. and Drissen, L.: 2014, \journal{Mon. Not. R. Astron. Soc.}, \vol{441}, 2996.

Arbutina, B., Uro{\v s}evi{\' c}, D., Stankovi{\' c}, M. and Te{\v s}i{\' c}, Lj.: 2004,
\journal{Mon. Not. R. Astron. Soc.}, \vol{350}, 346.

Arbutina, B. and Uro{\v s}evi{\' c}, D.: 2005, \journal{Mon. Not. R. Astron. Soc.}, \vol{360}, 76.

Blair, W. P., Sankrit, R. and Raymond, J. C.: 2005, \journal{Astrophys. J.}, \vol{129}, 2268.

Case, G. L. and Bhattacharya, D.: 1998, \journal{Astrophys. J.}, \vol{504}, 761.

Castelletti, G., Giacani, E., Dubner, G., Joshi, B. C., Rao, A. P. and Terrier, R.: 2011, \journal{Astron. Astrophys}, \vol{536}, 98.

Caswell, J. L., Haynes, R. F., Milne, D. K. and Wellington, K. J.: 1980, \journal{Mon. Not. R. Astron. Soc.}, \vol{190}, 881.

Chiotellis, A., Schure, K. M. and Vink, J.: 2012,  \journal{Astron. Astrophys}, \vol{537}, 139.

De Horta, A. Y., Collier, J. D., Filipovi{\'c}, M. D., Crawford, E. J., Uro{\v s}evi{\' c}, D.,
Stootman, F. H. and Tothill,  N. F. H.: 2013, \journal{Mon. Not. R. Astron. Soc.}, \vol{428}, 1980.

Dragicevich, P. M., Blair, D. G. and Burman, R. R.: 1999, \journal{Mon. Not. R. Astron. Soc.}, \vol{302}, 693.

Ferrand, G. and Safi-Harb, S.: 2012, \journal{Adv. Space Res.}, \vol{49}, 9, 1313.

Filipovi{\' c}, M. D., Payne, J. L. and Jones, P. A.: 2005, \journal{Serb. Astron. J.}, \vol{170}, 47.

Foster, T. J., Cooper, B., Reich, W., Kothes, R. and West, J.: 2013, \journal{Astron. Astrophys}, \vol{549}, A107.

Gerbrandt, S., Foster, T. J., Kothes, R., Geisb{\" u}sch, J. and Tung, A.: 2014,
\journal{Astron. Astrophys}, \vol{566}, A76.

Giacani, E. B., Dubner, G. M., Green, A. J., Goss, W. M. and Gaensler, B.M.: 2000, \journal{Astron. J.}, \vol{119}, 281.

Giacani, E., Smith, M. J. S., Dubner, G., Loiseau, N., Castelletti, G. and Paron, S.: 2009, \journal{Astron. Astrophys}, \vol{507}, 841.

Giacani, E., Smith, M. J. S., Dubner, G. and Loiseau, N.: 2011, \journal{Astron. Astrophys}, \vol{531}, 138.

Green, D. A.: 1984, \journal{Mon. Not. R. Astron. Soc.}, \vol{209}, 449.

Green, D. A.: 2004, \journal{Bull. Astr. Soc. India}, \vol{32}, 335.

Green, D. A.: 1991, \journal{Publ. Astron. Soc. Pac.}, \vol{103}, 209.

Green D. A.: 2014, 'A Catalogue of Galactic Supernova Remnants (2014 May version)',
Cavendish Laboratory, Cambridge, United Kingdom (available at "http://www.mrao.cam.ac.uk/surveys/snrs/").

Huang, Y.-L. and Thaddeus, P.: 1985, \journal{Astrophys. J.}, \vol{295}, L13.

Jeong, I-G., Koo, B-C., Cho, W.K., Kramer, C., Stutzki, J. and Byun, D-Y.: 2013, \journal{Astrophys. J.}, \vol{770}, 105.

Jiang, B., Chen, Y., and Wang, Q. D.: 2007, \journal{Astrophys. J.}, \vol{670}, 1142.

Jiang, B., Chen, Y., Wang, J., Su, Y., Zhou, Xin., Safi-Harb, S. and DeLaney, T.: 2010, \journal{Astrophys. J.}, \vol{712}, 1147.

Junkes, N., Fuerst, E. and Reich, W.: 1992, \journal{Astron. Astrophys. Suppl. Series}, \vol{96}, 1.

Kothes, R. and Foster, T.: 2012, \journal{Astrophys. J.}, \vol{746}, L4.

Leahy, D. A. and Tian, W. W.: 2007, \journal{Astron. Astrophys}, \vol{461}, 1013.

Leahy, D. A. and Green, K. S.: 2012, \journal{Astrophys. J.}, \vol{760}, 25.

Leahy, D., Green K. and Tian, W.: 2014, \journal{Mon. Not. R. Astron. Soc.}, \vol{438}, 1813.

Nikoli{\' c}, S., van de Vena, G., Hengb, K., Kupkoc, D., Husemannc, B. et al.: 2013, \journal{Science}, \vol{340}, 6128, 45.

Paron, S., Ortega, M. E., Petriella, A., Rubio, M., Dubner, G. and Giacani, E.: 2012, \journal{Astron. Astrophys}, \vol{547}, A60.

Paron, S., Weidmann, W., Ortega, M. E., Albacete Colombo, J. F. and Pichel, A.: 2013, \journal{Mon. Not. R. Astron. Soc.}, \vol{433}, 1619.

Pavlovi\'{c}, M. Z., Uro\v{s}evi\'{c}, D., Vukoti\'{c}, B., Arbutina, B. and G\"{o}ker, \"{U}. D.: 2013, \journal{Astrophys. J. Suppl. Ser.}, \vol{204}, 4. (Paper I)

Prinz, T. and Becker, W.: 2012, \journal{Astron. Astrophys}, \vol{544}, A7.

Prinz, T. and Becker, W.: 2013, \journal{Astron. Astrophys}, \vol{550}, A33.

Shklovskii, I. S.: 1960, \journal{Astron. Zh.} , \vol{37}, 256.

Su, H., Tian, W., Zhu, H. and Xiang, F. Y.: 2014, Proceedings of the International Astronomical Union, IAU Symposium, \vol{296}, 372.

Tian, W. W. and Leahy, D. A., 2012, \journal{Mon. Not. R. Astron. Soc.}, \vol{421}, 2593.

Tian, W. W. and Leahy, D. A., 2014, \journal{Astrophys. J.}, \vol{783}, L2.

Uchida, K., Morris, M. and Yusef-Zadeh, F.: 1992a, \journal{Astron. J.}, \vol{104}, 1533.

Uchida, K. I., Morris, M., Bally, J., Pound, M. and Yusef-Zadeh, F.: 1992b,  \journal{Astrophys. J.}, \vol{398}, 128.

Uchiyama, Y., Takahashi, T., Aharonian, F. A. and Mattox, J. R.: 2002, \journal{Astrophys. J.}, \vol{571}, 866.

Uro{\v s}evi{\' c}, D.: 2002, \journal{Serb. Astron. J.}, \vol{165}, 27.

Uro{\v s}evi{\' c}, D.: 2003, \journal{Astrophys. Space Sci.}, \vol{283}, 75.

Uro{\v s}evi{\' c}, D., Pannuti, T. G., Duric, N. and Theodorou, A.: 2005, \journal{Astron. Astrophys}, \vol{435}, 437.

Uro{\v s}evi{\' c}, D., Vukoti{\' c}, B., Arbutina, B. and Sarevska, M.: 2010,  \journal{Astrophys. J.}, \vol{719}, 950.

Vukoti{\' c}, B., Jurkovi{\' c}, M., Uro{\v s}evi{\' c}, D. and Arbutina, B.: 2014, \journal{Mon. Not. R. Astron. Soc.}, \vol{440}, 2026.

Xu, J. W., Han, J. L., Sun, X. H., Reich, W., Xiao, L., Reich, P. and Wielebinski, R.: 2007, \journal{Astron. Astrophys}, \vol{470}, 969.

Yamauchi, S., Nobukawa, M., Koyama, K. and Yonemori, M.: 2013, \journal{Publ. Astron. Soc. Jpn.}, \vol{65}, 6.

Yamauchi, S., Mimani, S., Ota, N. and Koyama, K.: 2014, \journal{Publ. Astron. Soc. Japan}, \vol{66}, 2.

Zhang, X., Chen, Y., Li, H. and Zhou, X.: 2013, \journal{Mon. Not. R. Astron. Soc.}, \vol{429}, L25.

Zhou, X., Chen, Y., Su, Y. and Yang, J.: 2009, \journal{Astrophys. J.}, \vol{691}, 516.

Zhu, H., Tian, W. W., Torres, D. F., Pedaletti, G. and Su,  H. Q.: 2013,  \journal{Astrophys. J.}, \vol{775}, 95.

Zhu, H. and Tian, W. W.: 2014, Proceedings of the International Astronomical Union, IAU Symposium, \vol{296}, 378.

Zhu, H., Tian, W. W. and Zuo, P.: 2014, \journal{Astrophys. J.}, \vol{793}, 95.

\endreferences

\end{multicols}

\vfill\eject

{\ }



\naslov{ A{\ZZ}URIRANA RELACIJA IZME{\DD}U POVR{\SS}INSKOG RADIO SJAJA I DIJAMETRA
ZA GALAKTICHKE OSTATKE SUPERNOVIH }


\authors{M. Z. Pavlovi{\'c}$^{1}$, A. Dobard{\v z}i{\' c}$^{1}$, B. Vukoti{\' c}$^{2}$ and D. Uro{\v s}evi{\' c}$^{1}$}
\vskip3mm



\address{$^1$Department of Astronomy, Faculty of Mathematics,
University of Belgrade\break Studentski trg 16, 11000 Belgrade,
Serbia}

\Email{marko@math.rs, aleksandra@math.rs, dejanu}{math.rs}

\address{$^2$Astronomical Observatory, Volgina 7, 11060 Belgrade 38, Serbia}

\Email{bvukotic}{aob.rs}

\vskip.7cm


\centerline{UDK \udc}


\centerline{\rit Originalni nauchni rad}

\vskip.7cm

\begin{multicols}{2}
{


{\rrm Predstavljamo a{\zz}uriranu empirijsku relaciju izme{\dd}u povr{\ss}inskog radio-sjaja i dijametra ($\Sigma-D$)
za ostatke supernovih u nashoj Galaksiji. Nash prvobitni kalibracioni uzorak Galaktichkih ostataka sa nezavisno
odre{\dd}enim daljinama je ponovo razmatran i dopunjen podacima koji su publikovani u prethodne dve godine.
Na kalibracioni uzorak u $\log \Sigma - \log D$ skali je primenjena metoda ortogonalnog fitovanja kao i metoda bazirana
na funkciji gustine verovatno{\cc}e.
Primenom nestandardne ortogonalne regresije postignuta je invarijantnost
relacija $\Sigma-D$ i $D-\Sigma$, u okviru intervala procenjene gre{\ss}ke.
Nashe prethodne Monte Karlo simulacije pokazale su da bi nagibi
empirijskih $\Sigma-D$ relacija trebalo da budu odre{\dd}ivani primenom ortogonalne regresije, koja se dobro pokazala
u primeni na uzorke sa znachajnim rasturanjem tachaka. Najnoviji uzorak kalibratora sadr{\zz}i 65 ljuskastih ostataka.
{\SS}est novih ostataka supernovih je dodato u uzorak iz rada Pavlovi{\cc}a i saradnika (2013, u da{\lj}em tekstu Chlanak \rm I), \rrm jedan je izostav{\lj}en i izmenjene su daljine za 10 ostataka.
Novi nagib je neznatno stmiji ($\beta \approx 5.2$) od nagiba $\Sigma-D$ relacije iz Chlanka \rm I \rrm ($\beta \approx 4.8$).
Metoda bazirana na funkciji gustine verovatno{\cc}e koristi mape
gustine koje omogu{\cc}avaju pouzdanija izrachunavanja i chuvaju vishe informacija sadr{\zz}anih u
kalibracionom uzorku. Izrachunali smo daljine do pet novih slabih galaktichkih ostataka supernovih otkrivenih
po prvi put od strane \rm Canadian Galactic Plane Survey \rrm i dobijene su daljine redom 2.3, 4.0, 1.3, 2.9 i 4.7 kiloparseka
za \rm G108.5+11.0, G128.5+2.6, G149.5+3.2, G150.8+3.8 i G160.1$-$1.1.
\rrm Koriste{\cc}i a{\zz}uriranu empirijsku relaciju, odredili smo daljine do
ljuskastih Galaktichkih ostataka i dobijeni rezultati me{\nj}aju njihovu skalu daljina i do 15 procenata, u odnosu
na Chlanak \rm I. \rrm Rachunanje iz funkcije gustine verovatno{\cc}e mo{\zz}e ponekad dati nekoliko puta ve{\cc}e ili manje
vrednosti u poredjenju sa vrednostima dobijenim ortogonalnim fitom ali u proseku ova razlika iznosi 32, 24 i 18 procenata za modu, medijanu
i srednju vrednost daljine.
}
}
\end{multicols}

\end{document}